\documentclass{article}

   \PassOptionsToPackage{numbers, compress}{natbib}

\usepackage[final]{neurips_2019}

\usepackage{graphicx}
\usepackage{float}

\usepackage{xcolor}

\usepackage{algorithmic}
\usepackage{algorithm}

\usepackage[utf8]{inputenc} 
\usepackage[T1]{fontenc}    
\usepackage{hyperref}       
\usepackage{url}            
\usepackage{booktabs}       
\usepackage{amsfonts}       
\usepackage{nicefrac}       
\usepackage{microtype}      
\usepackage{times}
\usepackage{helvet}
\usepackage{courier}
\usepackage[algo2e]{algorithm2e}
\usepackage{amsmath, bm}
\usepackage{amssymb}
\usepackage{amsthm}
\usepackage{amsfonts}

\SetKwInput{KwInput}{Input}
\SetKwInput{KwOutput}{Output}
\setcitestyle{square}

\title{Fairness in Multi-agent Reinforcement Learning for Stock Trading}

\author{Wenhang Bao\\
       Capgemini Invent,\\
       Email: wenhang.bao@capgemini.com   
       }

\begin{document}

\maketitle

\begin{abstract}


Unfair stock trading strategies have been shown to be one of the most negative perceptions that customers can have concerning trading and may result in long-term losses for a company. Investment banks usually place trading orders for multiple clients with the same target assets but different order sizes and diverse requirements such as time frame and risk aversion level, thereby total earning and individual earning cannot be optimized at the same time. Orders executed earlier would affect the market price level, so late execution usually means additional implementation cost. In this paper, we propose a novel scheme that utilizes multi-agent reinforcement learning systems to derive stock trading strategies for all clients which keep a balance between revenue and fairness. First, we demonstrate that Reinforcement learning (RL) is able to learn from experience and adapt the trading strategies to the complex market environment. Secondly, we show that the Multi-agent RL system allows developing trading strategies for all clients individually, thus optimizing individual revenue. Thirdly, we use the Generalized Gini Index (GGI) aggregation function to control the fairness level of the revenue across all clients. Lastly, we empirically demonstrate the superiority of the novel scheme in improving fairness meanwhile maintaining optimization of revenue. 
  
\end{abstract}

\section{Introduction}
\label{sect:introduction}

Developing the right stock trading strategies for the clients is a fundamental function of investment banks. Traditionally speaking, the ability to maximize trading revenue or minimize cost would be the main key performance indicator of these financial institutes, as it directly affects the profit of them. It has been said that, equality is of greater importance these days, as it influences customers' perceptions of fairness. As long as perceptions of fairness are used by people as a heuristic for trust, inequality may lead to a destruction of that trust. The execution of orders may lead to the feeling of unfairness. Investment banks and hedge funds have many clients with the same target assets but having different requirements including but not limited to order size and execution time frame. The execution of such orders is a time dependent behavior, as orders can not be executed at the same time, otherwise, it would cause significant price movement and implementation cost. However, if orders from clients are executed gradually, the early tradings would affect the implementation cost and risk of later tradings. A disagreement between the distribution of cost, risk and profit is known to affect the relationship between customers and sellers. As an instance, if customers believe that an investment bank prefers big volume orders so that these orders would be executed when there is low risk and high revenue, and their orders are relatively small, they would turn to other investment banks.

Fairness is a concept that has conventionally been studied in economics \cite{moulin2004fair} and political philosophy \cite{sep-rawls}. Recently, it has also become an important consideration in other applied fields, such as applied mathematics \cite{ogryczak2014fair} which focuses on solving fair optimization problems, machine learning \cite{busa2017multi,speicher2018unified,agarwal2018reductions,heidari2018fairness} whose data source could be highly-biased and artificial intelligence \cite{inbook} when investigating multi-agent systems. While an equal distribution of resources across all members within a group seems ideal, researchers have shown that inequalities are deemed appropriate \cite{van1999pursuit}, particularly when they optimize the outcome of the group. On the other hand, perceived inequalities have a strong impact on individuals' behavior, often motivating them to act contrarily to their rational self-interest to eliminate inequality \cite{camerer2011behavioral}. Although fairness is a key notion when dealing with multiple parties, it has only recently received attention in reinforcement learning. Also, fairness is only recently been emphasized by the financial industry, which pays attention to fairness in capital risk allocation, resource distribution and dynamic pricing.

In this paper, we propose to use multi-agent reinforcement learning (MARL) to resolve the fairness issue in stock trading. Reinforcement learning consists of agents interacting with the environment to learn an optimal policy by trial and error for sequential decision-making problems. It can develop optimal trading strategies in a stochastic stock trading market and optimize trading revenue. A multi-agent system means that we can build trading strategies for all clients individually. This is an important property as different clients have various requirements in stock trading, thus we cannot model different trading time frames, risk aversion level with single-agent reinforcement learning algorithm. Moreover, the multi-agent system improves the scalability of the system. Developing a trading strategy for one client with one RL agent is simple and convenient. Communication among these agents is anticipated to resolve the fairness issue related to stock trading. We use Generalized Gini Index aggregation function to measure the fairness level of the revenue across all clients.

The main contribution of this paper is the integration of fairness in the algorithm training process. We propose a novel multi-agent reinforcement learning training paradigm considering fairness level during the training process. First, we demonstrate that the algorithm can learn from experience and develop optimal trading strategies in a simulated trading environment. Secondly, we describe the way to integrate fairness in multi-agent RL algorithms. Agents would receive not only the environment observations but also the fairness metric. Thirdly, we empirically show that by taking into account fairness in the training process, the group would still achieve a comparable level of total utility, while keeping a balance between revenue and fairness. The total group utility, risk and cost are more equally distributed across all individuals within that group.

The remainder of this paper is organized as follows. Section \ref{sect:problem} describes the trading problem, especially the liquidation problem, reviews the Almgren-Chriss model that is used to simulate the market environment and formulate the concept of fairness. Section \ref{sect:drl} introduces the detailed settings of multi-agent reinforcement learning algorithm. Section \ref{sect:performance} presents the experimental results where we test our algorithms. Section \ref{sect:conclusion} concludes this paper and points out some future directions.  

\section{Problem Description}
\label{sect:problem}
In this section, we first describe the trading problem. To simplify the trading problem, we use the liquidation problem as an example of trading in this paper and explain why it is feasible to use reinforcement learning algorithms to address it. Then we describe the Almgren and Chriss model or the trading environment. Lastly, we introduce Generalized Gini Index which measures fairness level among clients. 

\subsection{Optimal Liquidation Problem}
\label{Liquidation}
We consider a trader who aims to sell $X$ shares of one stock within a time frame $T$. Liquidator's personal characteristics, such as risk aversion level $\lambda$, would remain unchanged throughout the process. The trader can either sell or not sell stocks, but cannot buy any stock during the time frame $T$. On the last day of the time frame, the liquidation process ends and the number of shares should be $0$. Since the trading volume is tremendous, the market price $P$ will drop during selling, temporarily or permanently, potentially resulting in enormous trading costs. 

The trader or the representative financial institute seeks to find an optimal selling strategy, minimizing the expected trading cost $E(X)$, or called \textit{implementation shortfall}, subject to certain optimization criterion. The trader would know all the environment information including price, historical price and number of trading days remaining. If there are $J$ traders, they would not know other traders' information. For instance, they would not know other traders' remaining shares or risk aversion levels.  

Based on the assumption that the trading would have market impacts as well as that agents and environment are interactive, it is feasible to train agents in the environment and derive liquidation strategies with reinforcement learning algorithms. 

\subsection{Environment Model for the Simulation}
\label{ACmodel}
The problem of an optimal liquidation strategy is investigated by using the Almgren-Chriss market impact model \cite{almgren2001optimal} on the background that the agents liquidate assets completely in a given time frame. The impact of the stock market is divided into three components: unaffected price process, permanent impact, and temporary impact. The stochastic component of the price process exists, but is eliminated from the mean-variance. The price process permits linear functions of permanent and temporary price. Therefore, the model serves as the trading environment such that when agents make selling decisions, the environment would return price information. 

The price process of the Almgren and Chriss model \cite{almgren2001optimal} is as follows:

\begin{itemize}
    \item Price under temporary and permanent impact
    $$
    P_k = P_{k-1} + \sigma \tau^{1/2} \xi_k -\tau g(\frac{n_k}{\tau}), k = 1,\ldots,N
    $$
where $\sigma$ represents the volatility of the stock, $\xi_k$ are random variables with zero mean and unit variance, $g(v)$ is a function of the average rate of the trading, $v = n_k/ \tau$ during time interval $t_{k-1}$ to $t_k$, $n_k$ is the number of shares to sell during time interval $t_{k-1}$ to $t_k$, $N$ is the total number of trades and $\tau = T/N$.
    \item Inventory process: $x_{t_k} = X - \sum_{j=1}^{k}n_j$, where $x_{t_k}$ is the number of shares remaining at time $t_k$, with $x_T = 0$.
    \item Linear permanent impact function
    $g(v) = \gamma v$, where $v = \frac{n_k}{\tau}$.
    \item Temporary impact function
    $h(\frac{n_k}{\tau}) = \epsilon~\text{sgn}(n_k) + \frac{\eta}{\tau} n_k$, where a reasonable estimate of $\epsilon$ is the fixed costs of selling, and $\eta$ depends on internal and transient aspects of the market micro-structure.
    \item Parameters $\sigma, \gamma, \eta, \epsilon$, time frame $T$, number of trades $N$ are set at $t = 0$.
\end{itemize}

\section{Deep Reinforcement Learning Algorithms}
\label{sect:drl}

We model the liquidation process as a Markov decision process (MDP), and then formulate the multi-agent setting used to resolve the problem. The training diagram is also covered, which explains how multiple agents interact and learn from environment in details. We use implementation shortfall as the metric of selling cost, and the properties of MDP process allows us to define the goal as minimizing the expected implementation shortfall.  

\subsection{Liquidation as a MDP Problem}
\label{DPL:liquidatoin}
 Considering the stochastic and interactive nature of the trading market, we model the stock trading process as a Markov decision process, which is specified as follows:
\begin{itemize}
    \item State $s = [\bm{r},m,l]$: a set that includes the information of the log-return $\bm{r} \in \mathbb{R}_+^D$, where $D$ is the number of days of log-return, and the remaining number of trades $m$ normalized by the total number of trades, the remaining number of shares $l$, normalized by the total number of shares. The log-returns capture information about stock prices before time $t_k$, where $k$ is the current step. It is important to note that in real world trading scenarios, this state vector may hold more variables. 
    \item Action $a$: we interpret the action $a_k$ as a selling fraction. In this case, the actions will take continuous values in between 0 and 1.
    \item Reward $R(s,a)$: to define the reward function, we use the difference between two consecutive utility functions. The utility function is given by:
    \begin{align}
        U(\bm{x}) &= E(\bm{x}) + \lambda V(\bm{x}) ,\label{eq:1}\\
        E(\bm{x}) &= \sum_{k=1}^{N}\tau x_k g(\frac{n_k}{\tau}) + \sum_{k=1}^{N}n_k h(\frac{n_k}{\tau}),\label{eq:2}\\
        V(\bm{x}) &= \sigma^2 \sum_{k=1}^{N}\tau x_k^2,\label{eq:3}
    \end{align}
        
where $\lambda$ is the risk aversion level, and $\bm{x}$ is the trading trajectory or the vector of shares remaining at each time step $k,~ 0 \le t_k \le T$. After each time step, we compute the utility using the equations for $E(\bm{x})$ and $V(\bm{x})$ from the Almgren and Chriss model for the remaining time and inventory while holding parameter $\lambda$ constant. Denoting the optimal trading trajectory computed at time $t$ by $\bm{x}_t^*$, we define the reward as:
        \begin{equation}
            R_{t} = {U_t(\bm{x}_t^*)-U_{t+1}(\bm{x}_{t+1}^*)} .
        \end{equation}
    \item Policy $\pi (s)$: The liquidation strategy of stocks at state $s$. It is essentially the distribution of selling percentage $a$ at state $s$.
    \item Action-value function $Q_\pi(s,a)$: the expected reward achieved by action $a$ at state $s$, following policy $\pi$.
\end{itemize}

\subsection{Multi-agent Reinforcement Learning Setting}
\label{DPL:Setting}
The advantage of multi-agent over single-agent reinforcement learning is the ability to incorporate high-level complexities in the system. The single-agent environment is a special case where the number of agents $J=1$. It simplifies the problem and would automatically inherit all properties from the multi-agent environment. Following the MDP configuration in the last section, we specify our multi-agent reinforcement learning setting as follows:

\begin{itemize}
    \item States $s = [\bm{r},m,\bm{l}]$ : in a multi-agent environment, the state vector should have information about the remaining stocks of each agent. Therefore, in a $J$ agents environment, the state vector at time $t_k$ would be:
        $$
        [r_{k-D},\ldots,r_{k-1},r_k,m_k,l_{1,k},\ldots,l_{J,k}],
        $$
where
    \begin{itemize}
        \item $r_k = \log(\frac{P_k}{P_{k-1}})$ is the log-return at time $t_k$.
        \item $m_k = \frac{N_k}{N}$ is the number of trades remaining at time $t_k$ normalized by the total number of trades.
        \item $l_{j,k} = \frac{x_{j,k}}{X_j}$ is the remaining number of shares for agent $j$ at time $t_k$ normalized by the total number of shares. 
    \end{itemize}

    \item Action $a$: using the interpretation in Section \ref{DPL:liquidatoin}, we can determine the number of shares to sell for each at each time step using:
        $$
        n_{j,k} = a_{j,k} \times x_{j,k} ,
        $$
where  $x_{j,k}$ is the number of remaining shares at time $t_k$ for agent $j$.

    \item Reward $R(s,a)$: denoting the optimal trading trajectory computed at time $t$ for agent $j$ by $x_{j,t}^*$, we define the reward as:
        \begin{equation}
            R_{j,t} = {U_{j,t}(\bm{x}_{j,t}^*)-U_{j,t+1}(\bm{x}_{j,t+1}^*)}.
        \end{equation}
        
    \item Observation $O$: Each agent only observes limited state information \cite{omidshafiei2017deep}. In other words, in addition to the environment information, each agent only knows its own remaining shares, but not other agents' remaining shares. The observation vector at time $t_k$ for agent $j$ is:
        $$
        O_{j,k} = [r_{k-D},\ldots,r_{k-1},r_k,m_k,l_{j,k}].
        $$
\end{itemize}

\subsection{Fairness}
Fairness can be defined in a theoretically-founded way \cite{moulin2004fair} and relies on two key principles. The first one (P1) is called “Equal treatment of equals”, which states that two users/stakeholders (with identical characteristics with respect to the optimization problem, as assumed in this paper) should be treated the same way. The second one (P2), called the Pigou-Dalton principle, is based on the notion of Pigou-Dalton transfer, which is a payoff transfer from a richer user/stakeholder to a poorer one without reversing their relative ranking. The Pigou-Dalton principle states that such transfers lead to more equitable distributions. Formally, for any $v \in \mathbb{R}^n$ where $v_i < v_j$ and for any $\epsilon \in (0,v_j-v_i)$ we prefer $v + \epsilon \text{1}_i$ $-$ $\epsilon \text{1}_j$ to $v$ where $\text{1}_i$ (resp. $\text{1}_j$ ) is the canonical vector, null everywhere except in component $i$ (resp. $j$) where it is equal to $1$. In other words, this principle states that, all other things being equal, we prefer more “balanced” distributions (i.e., vectors) of payoffs. Besides those two principles, as we are in an optimization context, an efficiency principle (P3) is also required, which states that given two payoff distributions, if one vector Pareto-dominates another, the former is preferred to the latter.

Those three principles imply that a fair welfare function that aggregates the payoffs of the users/stakeholders needs to satisfy three properties. They have to be symmetric (i.e., independent to the order of its arguments for P1), strictly Schurconcave (i.e., monotonic with respect to Pigou-Dalton transfers for P2) and strictly increasing (i.e., monotonic with respect to Pareto dominance for P3). The elementary approach based on maximin (or Egalitarian approach), where one aims at maximizing the worse-off user/stakeholder, does not satisfy the last two properties. A better approach is based on the lexicographic maximin, which consists in comparing first the worse-off user/stakeholder when comparing two vectors, then in the case of a tie, comparing the second worse-off and so on. However, due to the non-compensatory nature of the min operator, vector $(1, 1, . . . , 1)$ would be preferred to $(0, 100, . . . , 100)$, which may be debatable.

Many fair welfare functions have been proposed. In practice, the choice of a suitable function depends on the application domain. For illustration, we present the fair welfare function based on the Generalized Gini Index \cite{weymark1981generalized} $G_w: \mathbb{R}^n \to \mathbb{R}$:
\begin{equation}
    G_w(v) = \sum_i w_i v_i
\end{equation}
where $w \in [0,1]^n$ is a weight vector such as $w_1 > w_2 > \ldots >w_n$, and $(v_1,v_2,\ldots, v_n)$ is the payoff vector $v$ reordered in an increasing fashion. 

Function $G_w$ contains the welfare function induced by the classic Gini index or the Bonferroni index. We adjust the reward observed by the agents in our MARL algorithm by following updating function:

\begin{equation}
\label{eq:adjust}
    R_{j,t}:=R_{j,t}-w_j G_{w,t}(R_{\cdot,t}),
\end{equation}
where $R_{\cdot,t}$ is the reward vector received by all agents, $w$ is the weight vector computed by the percentage of initial shares of each agent to the total selling volume.

\subsection{Deep Reinforcement Learning Algorithm}
We adopt the Actor-Critic \cite{mnih2016asynchronous,lowe2017multi} method that uses neural networks to approximate both the Q-value and the action. The critic learns the Q-value function and uses it to update actor's policy parameters. The critic network estimates the expected return of a state-action pair.  
The actor brings the advantage of computing continuous actions without the need of a Q-value function, while the critic supplies the actor with knowledge of the performance. The actor network has state $s$ as input and returns action $a$ directly. Actor-critic methods usually have good convergence properties, in contrast to critic-only methods. 

\begin{algorithm}
    \caption{Multi-agent DDPG}
    \KwInput{number of episodes $M$, time frame $T$, minibatch size $N$, learning rate $\lambda$, and number of agents or intersections $J$}
    \begin{algorithmic}[1]
    \label{alg:DDPG}
        \FOR{$j = 1, J$}
            \STATE Randomly initialize critic network $Q_j(O_j,a|\theta_j^Q)$ and actor network $\mu_j(O_j|\theta_j^\mu)$ with random weight $\theta_j^Q$ and $\theta_j^\mu$ for agent $j$;
            \STATE Initialize target network $Q'_j$ and $\mu'_j$ with weights $\theta_j^{Q'} \leftarrow \theta_j^{Q}$, $\theta_j^{\mu'} \leftarrow \theta_j^{\mu}$ for each agent $j$;
            \STATE Initialize replay buffer $B_j$ for each agent $j$;
        \ENDFOR
        \FOR {episode $= 1, M$}
            \STATE Initialize a random process $\mathcal{N}$ for action exploration;
            \STATE Receive initial observation state $s_0$;
            \FOR{$t = 1, T$}
                \FOR{$j = 1, J$}
                    \STATE {Select action $a_{j,t} = \mu_j(O_{j,t}|\theta_j^\mu) + \mathcal{N}_t$ according to the current policy and exploration noise;}
                \ENDFOR
                \STATE {Each agent executes action $a_{j,t}$, market state changes to $s_{t+1}$;}
                \STATE {Each agent observes reward $r_{j,t}$ and observation $O_{j,t+1}$, where observation $O_{j,t+1}$ is adjusted according to spatial influence};
                \STATE {Each agent adjust reward according to Equation \ref{eq:adjust}}
                \FOR{$j = 1, J$}
                    \STATE {Store transition ($O_{j,t}$, $a_{j,t}$, $r_{j,t}$, $O_{j,t+1})$ in $B_j$;}
                    \STATE {Sample a random minibatch of $N$ transitions ($O_{j,i}$ , $a_{j,i}$ , $r_{j,i}$ , $O_{j,i+1}$) from $B_j$;}
                    \STATE {Set $y_{j,i} = r_{j,i}+\gamma Q'_j (s_{t+1}, \mu'_j (O_{j,i+1}|\theta_j^{\mu'}|\theta_j^{Q'}))$
                    for $i = 1, \ldots, N$;}
                    \STATE {Update the critic by minimizing the loss: $L = \frac{1}{N}\sum_i(y_{j,i} -Q_j(O_{j,i},a_{j,i}|\theta_j^Q))^2$;}
                    \STATE {Update the actor policy by using the sampled policy gradient:
                        \begin{multline*}
                            \nabla_{\theta^\mu} \pi \approx \frac{1}{N}\sum_i \nabla_a Q_j(O,a|\theta_j^Q)|_{O = O_{j,i},a = \mu_j(O_{j,i})}
                            \times \nabla_{\theta^\mu} \mu_j(O_j|\theta^\mu)|_{s_i};
                        \end{multline*}
                    }
                    \STATE {Update the target networks:\:\:$\theta_j^{Q'}\leftarrow \tau \theta_j^Q + (1-\tau)\theta_j^{Q'},\:\:\:\:\theta_j^{\mu'} \leftarrow \tau \theta_j^\mu + (1-\tau)\theta_j^{\mu'}.$}
                \ENDFOR
            \ENDFOR
        \ENDFOR
    \end{algorithmic}
\end{algorithm}

The Deep Deterministic Policy Gradients (DDPG) algorithm \cite{lillicrap2015continuous} is one example of an actor-critic method. We will use DDPG to generate the optimal execution strategy of liquidation. DDPG uses three skills to make sure that it gets converged experimental results: experience replay buffer, learning rate and exploration noise. Experienced replay method \cite{wang2016sample} enables the stochastic gradient decent method and removes correlations between consecutive transitions. Learning rate controls the updating speed of the neural network. Exploration noise addresses the exploration and exploitation trade-off. With these training skills, the agent would learn from trail and error and find the optimal trading trajectory that minimizes the trading cost. In other words, we will use the DDPG algorithm or Algorithm \ref{alg:DDPG} to solve the optimal liquidation problem. 

\section{Performance Evaluation}
\label{sect:performance}
We first describe the simulation environment in detail. We then use experimental results to demonstrate that for both plain DDPG algorithm and fairness integrated algorithm, the agents can learn optimal trading strategies. Moreover, for fairness integrated algorithm, the distributions of profit, cost and risk, which are measured by total utility, expected implementation shortfall and expected variance, respectively, are more evenly distributed within the group and fairly between groups. 

We implement a typical reinforcement learning workflow to train the actor and critic. We change the single-agent Almgren and Chriss model \cite{almgren2001optimal} settings to build the multi-agent environment. We adjust reward functions to integrate fairness component. We use Alg. \ref{alg:DDPG} to find a policy that can generate the optimal trading trajectory with minimum implementation shortfall. We feed the states observed from our simulator to each agent. These agents first predict actions using the actor model and perform these actions in the environment. Then, the environment returns their rewards and new states. This process continues for a given number of episodes. 

\subsection{Simulation Environment}
\label{Simulation_env}
This environment simulates stock prices that follow a discrete arithmetic random walk, and that the permanent and temporary market impact functions are linear functions of the rate of trading, as in the Almgren and Chriss model \cite{almgren2001optimal}. 

We set the initial stock price to be $P_0 = 50$. The stock price has $20\%$ annual volatility, a bid-ask spread of ${1}/{8}$, the difference between ask price and bid price, and an average daily trading volume of 50 million shares. Assuming that there are 250 trading days in a year, this gives a daily volatility in stock price of  ${0.12}/{\sqrt{250}}\approx 0.8\%$ . We use a liquidation time frame of $T = 60 $ days and we set the number of trades $N = 240$. This leads to $\tau = \frac{T}{N} = 4$ , which means that we will be making four trades per day. These settings are changeable and can be adjusted to same day liquidation as well. 

For the temporary cost function, we set the fixed cost of selling to be ${1}/{2}$ of the bid-ask spread, so $\epsilon = {1}/{16}$. We set $\eta$ such that for each one percent of the daily volume we trade, the price impact equals to the bid-ask spread. For example, trading at a rate of  $5\%$  of the daily trading volume incurs a one-time cost on each trade of ${5}/{8}$. Under this assumption we have  $\eta = {{(1/8)}/ {(0.01 \times 5\times 10^7)}}=2.5\times 10^7$.

For the permanent costs, a common rule of thumb is that price effects become significant when we sell $10\%$ of the daily volume. Here, by "significant" we mean that the price depression is one bid-ask spread, and that the effect is linear for both smaller and larger trading rates, then we have $\gamma = {{(1/8)}/{(0.1\times 5 \times 10^7)}} = 2.5 \times 10 ^8$.

\subsection{Experimental Results}

\begin{figure}
\begin{tabular}{cc}
\centering
  \includegraphics[width=56mm]{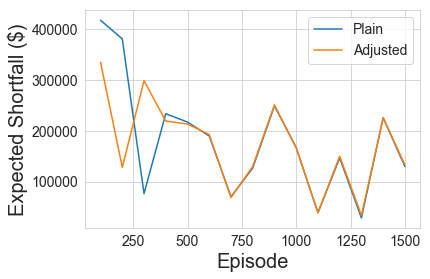} &   
  \includegraphics[width=50mm]{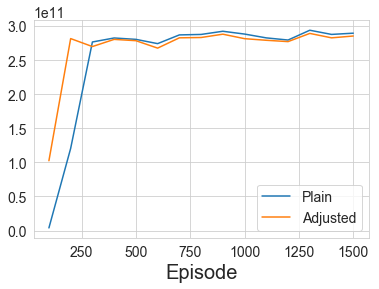}\\
(a) Implementation Shortfall & (b) Variance \\[6pt]
\end{tabular}
\caption{The total expected implementation shortfall (a) and total expected variance (b) for all clients converges to the same level for both algorithms.}
\label{fig:training}
\end{figure}

\begin{figure}
\begin{tabular}{cc}
\centering
  \includegraphics[width=52mm]{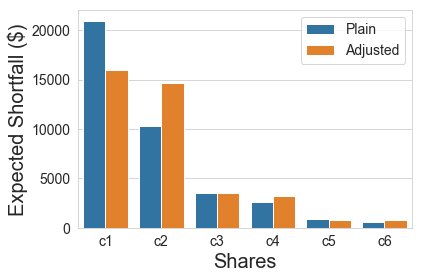} &   
  \includegraphics[width=50mm]{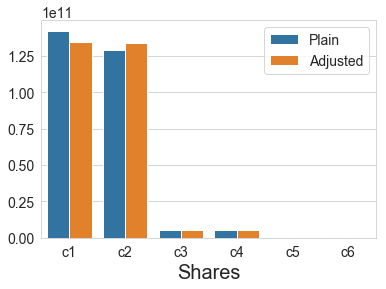}\\
(a) Implementation Shortfall & (b) Variance \\[6pt]
\end{tabular}
\caption{(a) The plain RL algorithm has larger inequalities both between groups and within the group. (b) Both algorithms have almost the same level of variance, which is a measure of risk. }
\label{fig:slice}
\end{figure}

We use 6 agents to develop trading strategies for 6 clients. They have 500k, 500k, 100k,100k,20k and 20k shares to sell, respectively. They have the same level of risk-aversion level $\lambda=1e-4$, and trading time frame $T=60$, etc. Their only difference is the trading volumes. We assess both the implementation shortfall and expected variance to demonstrate the superiority. The traditional multi-agent reinforcement learning algorithm, which we call plain algorithm, and fairness integrated multi-agent reinforcement learning algorithm are implemented to verify our expectations.

\subsubsection{Convergence}
Figure \ref{fig:training} shows that both algorithms have the ability to optimize the trading strategies, which implies that the adjustment of the reward function does not affect the capability of the algorithm to learn the optimal trading strategy. Moreover, the variance converges to the same level, which is an illustration that the adjustment does not affect the total risk as well.

Therefore, the adjustment of the RL algorithm does not affect either the total revenue or total risk of stock trading. It is important because if the adjustment changes the total implementation shortfall, it will affect the expected return of investment banks and hedge funds. Financial institutes are less likely to sacrifice revenue for fairness. Thus it builds the foundation for the pursuit of fairness for clients. 

\subsubsection{Fairness}
Figure \ref{fig:slice} shows that at the same level of total implementation shortfall, the adjusted RL algorithm has more evenly distributed profits. The profit level of clients from the same customers is closer in this case, which implies a fairer allocation of profit. Besides, we notice that it happens to variance as well. Therefore, for clients who have the same amount of stocks to sell, their implementation shortfall and risk would be similar, even though their orders are executed with sequence. 

In conclusion, the adjustment of reward function makes the distribution of profit, cost and risk fairer without sacrificing the total revenue and risk of the financial institutes. The fairness in profits ,cost and risk allocation helps both maintaining current clients and acquisition of new clients.

\section{Conclusion}
\label{sect:conclusion}
In this paper, we propose a novel scheme that integrated the fairness factor into the reinforcement learning algorithm. By using the Generalized Gini Index as a measure of fairness to adjust the reward in a multi-agent environment, agents can learn new liquidation policies that could achieve the same level of revenue while keeping a balance between revenue, cost and risk. The ability to maintain revenue is the main factor that makes the adjusted algorithm feasible and practical as it does not affect the earning capabilities of investment banks and hedge funds. Moreover, by considering the fairness factor in the training process, clients will be more satisfied with the service as long as the perceptions of fairness are used by clients as a heuristic for trust.

\end{document}